\documentclass[11pt]{article}

\usepackage[preprint]{acl}

\usepackage{times}
\usepackage{latexsym}
\usepackage[T1]{fontenc}
\usepackage[utf8]{inputenc}
\usepackage{microtype}

\usepackage{url}
\usepackage{graphicx}  
\usepackage{amsfonts}
\usepackage{CJK}
\usepackage{bm}
\usepackage{mathrsfs}
\usepackage{float}
\usepackage{xcolor,colortbl}
\usepackage{adjustbox}
\usepackage{subfigure}
\usepackage{booktabs,multirow}
\usepackage{bigstrut}
\usepackage{paralist}
\usepackage{bbm}
\usepackage{makecell}
\usepackage{nicefrac}       
\usepackage{microtype} 
\usepackage{epsfig}
\usepackage{diagbox}
\usepackage{framed}
\usepackage{empheq}
\usepackage{array}
\usepackage{enumitem}
\usepackage{mdwlist}
\usepackage{xspace,mfirstuc,tabulary}
\usepackage{algorithm}
\usepackage{placeins}
\usepackage{algpseudocode}
\usepackage{microtype}

\usepackage[normalem]{ulem}
\useunder{\uline}{\ul}{}
\usepackage{threeparttable}
\usepackage{makecell}
\usepackage{booktabs}

\usepackage{xspace}
\usepackage{natbib}
\usepackage{hyperref}
\usepackage{url}

\usepackage{graphicx}
\usepackage{subfigure}
\usepackage{multirow}
\usepackage{multicol}
\usepackage{pifont}
\usepackage{enumitem}
\usepackage{amsmath}
\usepackage{appendix}
\usepackage{fontawesome5}

\usepackage[figuresright]{rotating}
\usepackage[most]{tcolorbox}

\setlist[itemize]{nosep}

\newtcbox{\mybox}[1][red]
  {on line, arc = 0pt, outer arc = 0pt,
    colback = #1!10!white, colframe = #1!50!black,
    boxsep = 0pt, left = 1pt, right = 1pt, top = 2pt, bottom = 2pt,
    boxrule = 0pt, bottomrule = 1pt, toprule = 1pt}
\definecolor{BoxBackground}{RGB}{240, 240, 240} 
\definecolor{BoxFrame}{RGB}{0, 0, 0} 
\definecolor{TitleBackground}{RGB}{0, 0, 0} 
\definecolor{TitleText}{RGB}{255, 255, 255} 
\tcbset{
  academicbox/.style={
    boxsep=5pt,
    left=2pt,
    right=2pt,
    bottom=0.5pt,
    boxrule=0.5pt,
    colback=BoxBackground,
    colframe=BoxFrame,
    colbacktitle=TitleBackground,
    coltitle=TitleText,
    enhanced,
    attach boxed title to top left={yshift=-0.1in,xshift=0.1in},
    boxed title style={boxrule=0pt,colframe=white},
    title={#1},
  }
}

\title{Intent$^\uparrow$Coding: Amplifying User Intent in Code Generation}
\author{
Zheng Fang\textsuperscript{1},
Yihong Dong\textsuperscript{1},
Lili Mou\textsuperscript{2,3},
Dongming Jin\textsuperscript{1},
Zhi Jin\textsuperscript{1},
Ge Li\textsuperscript{1}
\\
\textsuperscript{1} School of Computer Science, Peking University \\
\textsuperscript{2} Department of Computing Science, University of Alberta \\
\textsuperscript{3}  Canada CIFAR AI Chair \\
\texttt{fangz@pku.edu.cn} \quad
\texttt{dongyh@stu.pku.edu.cn} \quad
\texttt{lige@pku.edu.cn}
}

\begin{document}
\maketitle


\renewcommand{\thefootnote}{\arabic{footnote}}

\begin{abstract}

\newcommand{\model}{\textit{Intent}$^\uparrow$\!\textit{Coding}}
\newcommand{\data}{\textit{CodeConstraints}}
Large Language Models (LLMs) have shown strong capabilities in code generation, but their adherence to fine-grained user intent with multiple constraints remains a significant challenge. Our empirical analysis reveals two key observations: 1) Model performance deteriorates quickly as the number of constraints in the user intent increases, and 2) While user intent does influence the model's logits, such an influence may not be strong enough to effectively steer the decoding process.
To this end, we propose Intent-Amplified Code Generation (\model), a novel decoding strategy that enhances an LLM's ability to follow user intent. \model{} captures the influence of user intent by masking out the intent, and applies a multi-strength ensemble mechanism to amplify the effect of user intent during generation. \model{} is model-agnostic, requires no additional training, and integrates seamlessly with existing decoding procedures. To enable systematic evaluation, we also construct \data{}, a benchmark dataset specifically designed to test user intent compliance under varying numbers of constraints. Experiments on our constructed \mbox{\data}, as well as popular IFEvalCode, HumanEval and LiveCodeBench datasets, show that our \model{} model significantly improves both constraint satisfaction and functional correctness compared to standard decoding approaches. \model{} achieves up to 71.0\% relative improvement on \data{}, achieves up to 67.3\% relative improvement on IFEvalCode and achieves up to 29.3\% relative improvement in $\mathrm{pass@1}$ on HumanEval and LiveCodeBench compared with greedy decoding. 

\end{abstract}

\section{Introduction} \label{sec:intro}
Code generation, a fundamental task in software engineering, refers to the automatic creation of programming code from a certain form of user intent, typically given by a natural language description~\cite{DBLP:conf/sigsoft/SvyatkovskiyDFS20,DBLP:journals/corr/abs-2107-03374,li2022competition}. Code generation can significantly accelerate software development and reduce human efforts~\cite{DBLP:journals/tosem/DongJJL24,DBLP:journals/corr/abs-2406-00515}. The recent success of large language models (LLMs) has catalyzed a new paradigm in this area, where LLMs are trained on large amounts of data to generate syntactically and semantically correct code~\cite{li2022competition,codellama}. This paradigm is particularly promising for improving developers' productivity and enabling novice programmers to accomplish complex tasks.

\begin{figure}[tb]        
  \centering
  \includegraphics[width=\columnwidth]{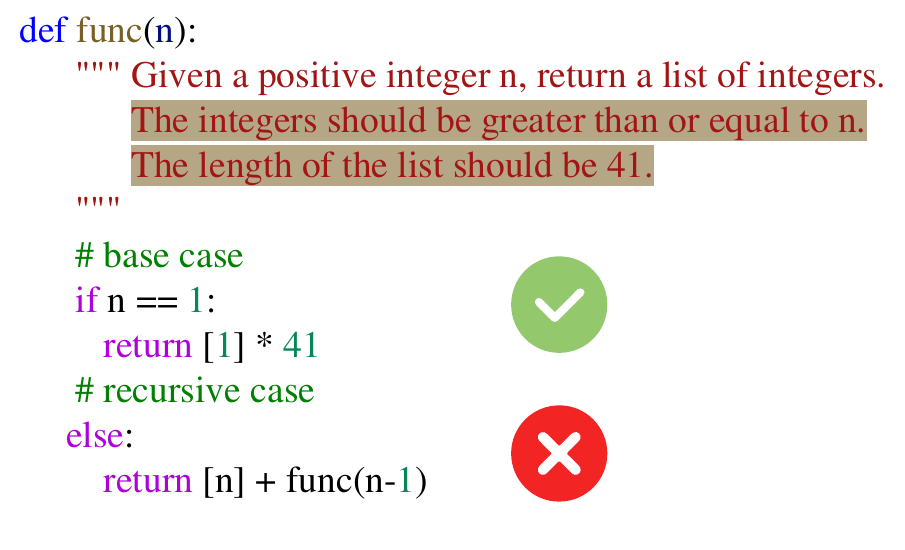}
  \caption{An example of generated code given by CodeLlama~\cite{codellama}, where the highlighted constraint is not satisfied. CodeLlama successfully handles an edge case ($n=1$, green \faCheckCircle). But the code (red \faTimesCircle) violates the required constraints when considering a broader range of inputs.}
  \label{fig:example}
\end{figure}

Previous work on LLM-based code generation is typically based on autoregressive models, where each token is predicted based on the user intent (natural language description) as well as all previously generated tokens~\cite{DBLP:journals/tosem/JiangDWFSLJJ24}. 
To translate the model's probabilistic outputs into final code, researchers have explored various decoding methods, e.g., greedy decoding and sampling methods~\cite{DBLP:conf/aclnmt/FreitagA17,DBLP:conf/aaai/ZhuLLZLJ024}. 

However, a user intent may specify a number of constraints for a desired program, but a standard decoding method often fails to fully satisfy the constraints. In Figure~\ref{fig:example}, for example, the natural language descriptions specifies that 1) The code should return a list, 2) The element in list should be integer, 3) The integer in list should be greater than or equal to $n$, and 4) The length of the list should be 41. However, the code generated by CodeLlama~\cite{codellama} satisfies constraints 1) and 2), but fails to satisfy 3) and 4).

In our work, we make two key observations on how an LLM addresses, or fails to address, the user intent. \textbf{Observation 1:}  LLM's performance deteriorates quickly as the number of user-specified constraints increases. To interpret the phenomenon, we compare the logits with and without feeding the user intent, leading to \textbf{Observation 2:} Although feeding the user intent influences the LLM's logits, this effect is often too weak to guide standard decoding methods effectively, resulting in failures to satisfy certain constraints.

\newcommand{\model}{\textit{Intent}$^\uparrow$\!\textit{Coding}}
\newcommand{\data}{\textit{CodeConstraints}}

To this end, we propose Intent-Amplified Code Generation (\model{}), a novel decoding strategy that enhances an LLM's compliance with user intent. At each decoding step, our method generates the original logits using a prefix of the user intent and previously generated tokens, while simultaneously computing intent-masked logits using the same prefix with the user intent masked. Contrasting the original logits and the intent-masked logits captures the influence of user intent. The influence is then scaled and added to the original logits. Our approach is inspired by contrastive decoding~\cite{DBLP:conf/acl/LiHFLEHZL23,DBLP:conf/iclr/KimKLY24}, but differs in that we do not rely on a single, fixed scale of the contrastive signal. Instead, our method explores multiple scale values at each step and ensembles them during beam search. This allows the decoding process to dynamically adjust the contrastive scale for amplifying user intent. Notably, our method is model-agnostic and training-free. 

To facilitate research on multi-constraint user intent modeling, we construct a new benchmark dataset, called \data. We accomplish this by combining one or more primitive constraints (e.g., data type, return type, length, and value) to form a user intent. We control the task difficulty by varying the number of active constraints. Compared with existing datasets, our constructed \data{} have unique features: each primitive constraint is simple and within the capability of current LLMs. Consequently, our dataset focuses on testing LLMs' composability of multiple constraints in user intent. Further, our methodology of data construction is highly extensible and flexible, and can be easily scaled to arbitrary sizes and adapted to include new constraint types.

We evaluated our \model{} approach on two general benchmarks HumanEval~\cite{DBLP:journals/corr/abs-2107-03374} and LiveCodeBench~\cite{DBLP:conf/iclr/JainHGLYZWSSS25}, as well as two constraint-following benchmarks, IFEvalCode~\cite{ifevalcode} dataset and our constructed \data{} dataset. The experimental results consistently demonstrate the effectiveness of our approach.

In summary, our contributions are as follows: 1) We propose a novel approach, Intent Amplified Code Generation (\mbox{\model}), that enhances the model's ability to follow user intent. 2)  We construct \data, a new benchmark dataset specifically designed to evaluate how well LLMs can satisfy user intent with multiple constraints. 3) We conduct comprehensive experiments on both general and proposed benchmarks, demonstrating the effectiveness and superiority of our method.



\section{Related Work} \label{sec:related_works}

\subsection{Code Generation with LLMs}
Recent advances in LLMs have driven substantial progress in code generation. Foundational models such as AlphaCode~\cite{li2022competition} and CodeGen~\cite{DBLP:conf/iclr/NijkampPHTWZSX23} demonstrated the viability of this approach for complex programming tasks. This has been followed by a wave of powerful open-source models, including StarCoder~\cite{starcoder}, CodeLlama~\cite{codellama}, DeepSeek-Coder~\cite{deepseek}, and Qwen2.5-Coder~\cite{qwen}, which have further advanced the state of the art.

Beyond scaling base models, another line of research focuses on enhancing code generation at the decoding stage of LLMs. Execution-guided methods leverage runtime feedback to improve code quality. Some of these approaches generate multiple candidate programs and re-rank them based on test case outcomes~\cite{li2022competition,self-debug}. Others employ more sophisticated techniques like Monte Carlo Tree Search to explore promising solution paths informed by execution results~\cite{DBLP:conf/iclr/ZhangCSDTG23}. To reduce reliance on sampling, recent methods employ iterative refinement: Self-Edit~\cite{DBLP:conf/acl/ZhangLLLJ23} introduces an auxiliary model that corrects erroneous outputs based on test feedback, while ROCODE~\cite{rocode} integrates backtracking and static analysis into the decoding process, enabling the model to detect and repair errors during generation. Although effective, these techniques often depend on external feedback, such as test cases or compilers, limiting their applicability in real-world settings where such signals are unavailable.

Code generation benchmarks have evolved from basic function synthesis to more realistic and context-rich tasks. HumanEval~\cite{DBLP:journals/corr/abs-2107-03374} and MBPP~\cite{mbpp} evaluate models on standalone Python functions with unit tests. APPS~\cite{apps} scales up to thousands of real-world problems, emphasizing broader reasoning and domain coverage. LiveCodeBench~\cite{DBLP:conf/iclr/JainHGLYZWSSS25} is a continuously updated benchmark that assesses LLMs on recent competitive programming problems from online platforms. These benchmarks enable comprehensive and reliable evaluation of code generation capabilities across diverse scenarios.

\subsection{Decoding Methods}
Contemporary LLM-based code generation relies fundamentally on decoding strategies to balance precision and expressive flexibility. Traditional techniques can be categorized into deterministic and sampling‑based approaches. Greedy decoding selects the single highest‑probability token at each time step, while beam search~\cite{DBLP:conf/aclnmt/FreitagA17} maintains multiple top‑scoring hypotheses to produce a high-probability sequence in a global sense. Sampling‑based methods generate the token in a stochastic way, but often the distribution is reshaped before selecting tokens: temperature sampling~\cite{temperature} stretches or sharpens logits to modulate randomness; top‑k sampling~\cite{topk} restricts choices to the k most probable tokens; and nucleus sampling~\cite{topp} dynamically includes all tokens whose cumulative probability exceeds a threshold $p$, followed by renormalization prior to stochastic selection. Building on temperature sampling, AdapT sampling~\cite{DBLP:conf/aaai/ZhuLLZLJ024} analyzes token‑level loss distributions, classifies tokens into challenging tokens (e.g., block‑start) and confident tokens, and applies a higher temperature to challenging tokens and a lower temperature for confident tokens. Selective Prompt Anchoring (SPA) ~\cite{spa} adjusts the next-token logits to emphasize user-specified context. But its effectiveness depends on tuning a hyperparameter on a calibration dataset split. By contrast, our method does not need this kind of hyperparameter search for a single best strength.
\section{Methodology} \label{sec:approach}
\subsection{Construction of the \data{}}
Researchers have proposed a variety of benchmarks such as HumanEval~\cite{DBLP:journals/corr/abs-2107-03374} and LiveCodeBench~\cite{DBLP:conf/iclr/JainHGLYZWSSS25} to evaluate code generation. While these benchmarks cover diverse tasks and granularities, they primarily assess functional correctness through test case execution. However, some prompts in these benchmarks contain ambiguous user intent~\cite{DBLP:conf/scam/SiddiqDSS24}, and some include overly complex user intent that exceed the capabilities of current LLMs~\cite{DBLP:journals/corr/abs-2405-11430}. These benchmarks also lack the ability to assess whether specific constraints are satisfied when multiple constraints are present in the prompts. This gap motivates the design of our benchmark.

In our work, we constructed \data{}, a new benchmark dataset for code generation that emphasizes the capture of user intent. The dataset provides a controlled environment where the complexity and quantity of constraints can be precisely manipulated.

The data construction process is founded on four core primitive constraints:
\begin{itemize}
    \item \textbf{Data Type:} Specifying the numerical type of the elements to be generated (e.g., \texttt{integer}, \texttt{float}).
    \item \textbf{Return Format:} Defining the collection type for the output (e.g., \texttt{list}, \texttt{tuple}, \texttt{set}).
    \item \textbf{Length:} Imposing a constraint on the exact number of elements in the returned collection.
    \item \textbf{Value:} Restricting the numerical range of the elements (e.g., must be greater than or less than a given value).
\end{itemize}

These primitives are combinatorially sampled to construct problems with varying levels of difficulty, where the level number corresponds to the quantity of active constraints in the prompt. For instance, a Level 1 task might only specify the return format, whereas a Level 4 task combines all four primitive types into a single user intent. This hierarchical structure is designed to enable a granular analysis of model performance as the number of constraints increases.

We generate the \data{} dataset programmatically using a set of pre-defined constraint rules. This rule-based approach offers significant advantages. It ensures both scalability and extensibility. The dataset is extensible to arbitrary sizes. Our benchmark consists of 300 Level 2 problems, 100 Level 3 problems, and 100 Level 4 problems. Level 1 problems are trivial and can be easily solved by all major LLMs. Its modular design also allows new constraint types to be easily integrated for future work. Despite this potential for expansion, our experiments confirm the current dataset is already highly challenging, proving sufficiently difficult for current LLMs.

For evaluation, we use the accuracy metric, where a program is considered correct if all constraints are satisfied. This follows the rigid nature of program functionality, as opposed to more flexible natural language.




\subsection{Empirical Observations}

\begin{figure}        
  \centering
  \includegraphics[width=\columnwidth]{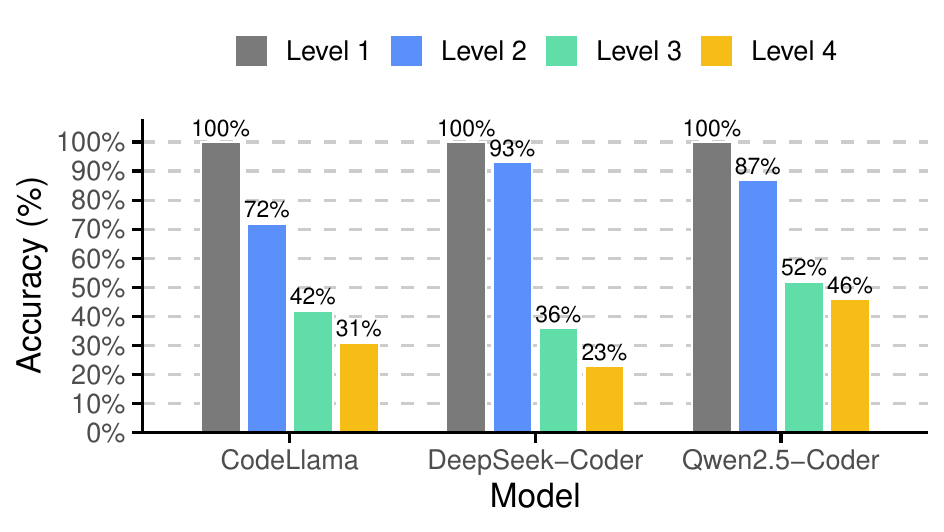}
  \caption{Performance of the 7B models on the \data{} dataset across various complexity levels, using greedy search for all evaluations.}
  \label{fig:icr}
\end{figure}

\begin{figure}
    \centering
    \includegraphics[width=0.99\linewidth]{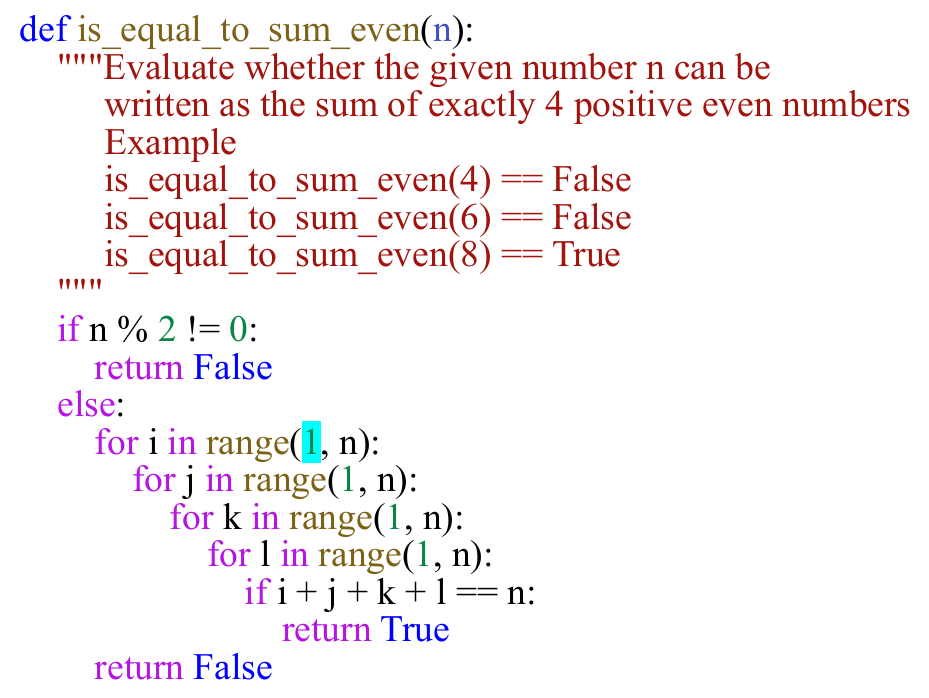} \\[0.5em]
    \includegraphics[width=0.99\linewidth]{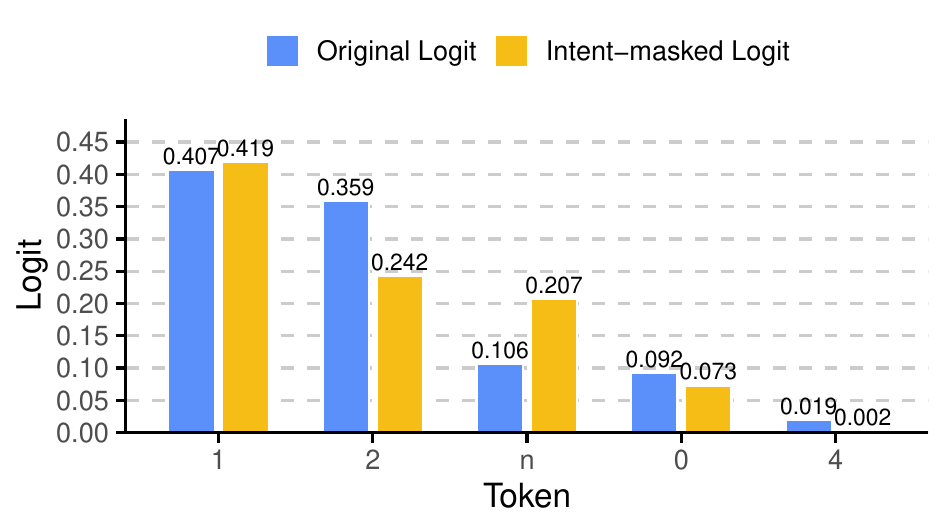} 
    \caption{An example of generated code given by DeepSeek-Coder~\cite{deepseek}, where the highlighted token is the wrong token and the top-5 token logits for this position are shown below.}
    \label{fig:ob2}
\end{figure}

We conduct an empirical analysis on the failure of LLMs to adhere to user intent during code generation. Our analysis leverages the HumanEval and our \data{} benchmarks. This investigation yields two key observations.


\paragraph{Observation 1.} Our first observation is that the performance of LLMs on code generation deteriorates quickly as the number of user-specified constraints increases. This trend is strongly supported by empirical evidence from our \data{} benchmark. As shown in figure~\ref{fig:icr}, DeepSeek-Coder drops from 93\% accuracy with two constraints to 23\% with four. Similar declines are observed for other LLMs. The high performance under low-constraint settings indicate that LLMs are capable of understanding and satisfying individual constraints. However, their compliance ability degrades significantly as task complexity increases.

\begin{figure*}        
  \centering
  \includegraphics[width=\textwidth]{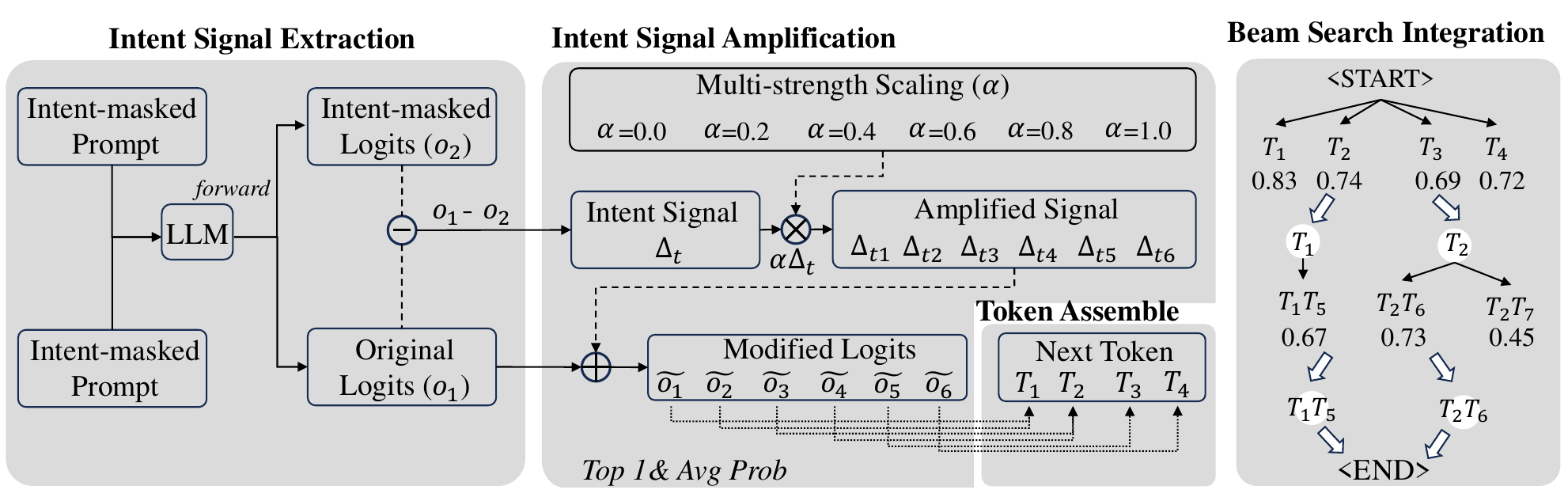}
  \caption{\textbf{Overview of \model{}.} We feed the original prompt and intent-masked prompt to the LLM in a same batch (details in Appendix~\ref{app:prompt}). We extract the intent signal, scale it with multiple strengths $\alpha$, and apply a token-level ensemble to obtain a set of candidate next tokens. These candidates are then used to expand each beam hypothesis; the right panel illustrates beam size $=2$.}
  \label{fig:method}
\end{figure*}

\paragraph{Observation 2.} Our second observation is that user intent may fail to sufficiently impact the LLMs' logits during decoding, even when the intent appears to be understood. As shown in Figure~\ref{fig:ob2}, we analyze a case where DeepSeek-Coder is asked to generate a function that determines whether an integer $n$ can be written as the sum of four positive even numbers. The LLM initially shows correct understanding by checking whether $n$ is even (\texttt{if n \% 2 != 0:}). But it generates incorrect code \texttt{for i in range(1, n):} as it ignores the ``even number'' constraint in the user intent. We analyze the original logits (with the user intent fed) and the intent-masked logit (with the user intent masked out). We see that, if we feed the user intent to the LLM, the logit for the correct token \texttt{2} increases from $0.242$ to $0.359$. While the logit for the incorrect token \texttt{1} decreases from $0.419$ to $0.407$. However, the logit for \texttt{1} remains higher and under greedy decoding, the LLM still selects the incorrect token. This indicates that while the model incorporates some aspects of the intent, the influence is insufficient to overcome the default tendency, leading to constraint non-compliance.


\subsection{The Proposed \model{} Method}

Based on our observations, we propose Intent-Amplified Code Generation (\model{}), a novel decoding strategy designed to improve LLMs' compliance with user intent. Figure~\ref{fig:method} provides an overview of the \model{} pipeline. The core idea is to run an LLM with and without the user intent separately. We enhance user intent in a contrastive way based on the intent-masked logits. Different from traditional contrastive decoding~\cite{DBLP:conf/acl/LiHFLEHZL23,DBLP:conf/iclr/KimKLY24, spa}, however, we do not use a fixed contrastive strength but build an ensemble of multiple strengths. Overall, our \model{} consists of four main stages.

\paragraph{Extracting the Intent Signal.} 
As demonstrated in Observation 2, user intent can subtly influence the LLMs' logit distribution during decoding, yet this influence is often too weak to determine the final output. To capture this intent signal, we construct an intent-masked prompt from the original prompt by masking attention of the user intent. This yields two sets of logits at each decoding step $t$: the original logit $o_t(\cdot| \text{prompt}_{\text{orig}}, x_{<t})$ and the intent-masked logit $o_t(\cdot| \text{prompt}_{\text{masked}}, x_{<t})$, where $\cdot$ represents a token in the vocabulary. The difference between the logits quantifies the influence of the user intent on the LLM's token prediction at each step:
\begin{equation}
\begin{aligned}
\Delta_t(\cdot) = o_t(\cdot| \text{prompt}_{\text{orig}}, x_{<t}) \\
- o_t(\cdot| \text{prompt}_{\text{masked}}, x_{<t})
\end{aligned}
\end{equation}

\paragraph{Amplifying the Signal.} 
To amplify the influence of user intent, we modify the LLM's logits with a scaled intent signal: 
\begin{equation}
\tilde{o_t}(\cdot) = o_t(\cdot|\text{prompt}_{\text{orig}}, x_{<t}) + \alpha \Delta_t(\cdot)
\label{eq:alpha}
\end{equation}
The choice of the scaling factor $\alpha$ is critical. A small $\alpha$ may fail to sufficiently amplify the influence, while a large $\alpha$ can distort the LLM's original distribution and introduce undesired bias. To balance this, we select six evenly spaced values from the interval $[0,1]$:
$$A = [0, 0.2, 0.4, 0.6, 0.8, 1.0]$$
where $ \alpha = 0 $ corresponds to no modification, and $ \alpha = 1 $ applies the full strength of the intent signal. We will build an ensemble of different values, so our \model{} approach does not rely on a fixed hyperparameter.

\paragraph{Token-Level Ensemble.}\label{para:tok}
For each $\alpha \in A$, we compute the logit based on Eqn.~(\ref{eq:alpha}) and select the top-1 token with the highest probability, yielding a set of candidate tokens. We then perform an ensemble over these tokens: for each unique token, we average its softmax probabilities across all the distributions where it was selected. More implementation details and an example are provided in Appendix~\ref{app:tok}.
This ensemble mechanism aggregates evidence across multiple influence strengths, favoring tokens that are robustly supported under varying degrees of intent amplification. 
\begin{table*}[t]
    \centering
    \small
    \setlength{\tabcolsep}{6pt}
    {\begin{tabular}{lllllll}
        \toprule
        \textbf{Model} & \textbf{Method} & \textbf{CodeConstraints} & \textbf{IFEvalCode} & \textbf{HumanEval} & \textbf{LiveCodeBench} & \textbf{Average} \\ 
        \midrule
        \multirow{8}{*}{CodeLlama}      & Greedy          & 31.0                    & 19.0                & 31.1               & 6.8                    & 13.0                 \\
               & Beam\_Search      & 18.0                    & 15.2                & 34.2               & 6.6                    &  11.9                \\
               & NS~($p=0.7$)         & 35.7$\pm1.2$                    & 18.7$\pm0.4$                & 31.3$\pm1.2$               & 7.4$\pm0.1$                    &    13.8              \\
               & NS~($p=0.8$)         & 33.7$\pm1.2$                    & 16.2$\pm0.8$                & 31.5$\pm2.2$               & 7.1$\pm0.1$                    &    13.2              \\
               & NS~($p=0.9$)         & 33.7$\pm1.2$                    & 15.9$\pm1.2$                & 30.9$\pm1.7$               & 7.0$\pm0.2$                    &  13.0                \\
               & AdapT           & 31.3$\pm1.2$                    & 18.9$\pm1.1$                & 34.2$\pm0.9$               & \textbf{7.5}$\pm0.1$                    &   13.9               \\
               & SPA             & \textbf{44.0}                    &  \textbf{21.9}                   & \textbf{37.2}               & 6.3                    &   \textbf{14.7}               \\
               & \model{}    & \textbf{53.0}$_{+71.0\%}$                    &  \textbf{24.8}$_{+30.5\%}$                   & \textbf{40.2}$_{\text{+29.3}\%}$               & \textbf{8.4}$_{\text{+23.5}\%}$    & \textbf{17.5}$_{\text{+34.6}\%}$                 \\ \midrule
        \multirow{8}{*}{Deepseek‑Coder}       & Greedy          & 23.0                    & 17.1                & 48.2               & 14.9                   &   20.1               \\
               & Beam\_Search      & 28.0                    & 23.8                & 51.8               & \textbf{16.4}                   &  \textbf{22.6}                \\
               & NS~($p=0.7$)         & 21.0$\pm2.4$                   & 17.1$\pm1.1$                & 46.7$\pm0.3$               & 14.5$\pm0.4$                   &  19.5                \\
               & NS~($p=0.8$)         & 19.3$\pm1.2$                    & 16.8$\pm1.2$                & 43.5$\pm1.5$               & 14.4$\pm0.4$                   &  18.8                \\
               & NS~($p=0.9$)         & 19.7$\pm1.2$                    & 14.9$\pm1.9$                & 42.4$\pm1.2$               & 14.3$\pm0.5$                   &   18.5               \\
               & AdapT           & 20.3$\pm1.9$                    & 17.0$\pm1.6$                & 45.3$\pm0.6$               & 14.8$\pm0.3$                   &      19.4            \\
               & SPA             & \textbf{29.0}                    & \textbf{24.8}                & \textbf{53.7}               & 15.6                   &           22.4       \\ 
               & \model{}    & \textbf{39.0}$_{\text{+70.0}\%}$                   & \textbf{28.6}$_{\text{+67.3}\%}$                & \textbf{59.8}$_{\text{+24.1}\%}$               & \textbf{17.8}$_{\text{+19.5}\%}$         &   \textbf{25.9}$_{\text{+28.9}\%}$               \\ \midrule
        \multirow{8}{*}{Qwen2.5‑Coder}          & Greedy          & 46.0                    & 23.8                & 60.4               & 25.1                   &     31.3             \\
               & Beam\_Search      & 35.0                    & 27.6                & \textbf{70.1}               & \textbf{27.7}                   &  \textbf{33.8}                \\
               & NS~($p=0.7$)         & 47.0$\pm2.9$                    & 24.8$\pm0.8$                & 59.2$\pm0.1$               & 23.9$\pm0.1$                   &   30.5               \\
               & NS~($p=0.8$)        & 46.0$\pm1.4$                    & 26.0$\pm0.5$                & 56.9$\pm0.2$               & 24.2$\pm0.5$                   &     30.4             \\
               & NS~($p=0.9$)         & 45.3$\pm0.9$                    & 25.1$\pm1.2$                & 56.1$\pm0.1$               & 23.8$\pm0.2$                   &   29.9               \\
               & AdapT           & 43.3$\pm0.5$                    & 26.7$\pm1.8$                & 60.6$\pm2.3$               & 24.9$\pm0.2$                   &     31.2             \\
               & SPA             & \textbf{69.0}                    & \textbf{28.6}                & 64.0               & 24.2                   &     33.4             \\
               & \model{}    & \textbf{65.0}$_{\text{+41.3}\%}$                    & \textbf{33.3}$_{\text{+40.0}\%}$                & \textbf{65.2}$_{\text{+7.9}\%}$               & \textbf{28.5}$_{\text{+13.5}\%}$       & \textbf{36.6}$_{\text{+16.9}\%}$ \\ 
        \bottomrule
    \end{tabular}
    }
    \caption{The main results of our experiments. \textbf{NS} denotes Nucleus Sampling~\cite{topp}. \textbf{SPA} denotes Select Prompt Anchoring~\cite{spa}. All sampling-based methods are run three times, and we report the mean and standard deviation. The subscript values represent the relative improvement of \model{} over the greedy search baseline.The top-2 results among all methods are \textbf{bold-faced}.}
    \label{tab:main_result}
\end{table*}
\paragraph{Beam Search Integration.} While sampling-based decoding is commonly used for text generation, beam search remains the standard decoding method for code generation, which is a more deterministic task than open-ended text generation~\cite{DBLP:conf/acl/IppolitoKSKC19,DBLP:conf/emnlp/ShiY0ZWYL24}.
We integrate our token-level ensemble strategy into a beam search decoding process to enable a more robust search. At each decoding step, we use the candidate tokens, obtained in the previous stage, to expand the current beam hypotheses. This process creates a diverse set of new hypotheses, each reflecting a distinct path under varying degrees of intent amplification. The expanded set of hypotheses is then pruned to the beam size by retaining only those with the highest cumulative log-probabilities. This integration allows the search to actively explore multiple intent-amplified paths, ultimately selecting code that balances overall fluency with precise adherence to user intent.

\section{Experiments} \label{sec:evaluation}

\subsection{Experimental Setup}
We evaluate constraint following ability on our \mbox{\data} and IFEvalCode~\cite{ifevalcode} dataset, and assess general code generation ability on HumanEval~\cite{DBLP:journals/corr/abs-2107-03374} and LiveCodeBench~\cite{DBLP:conf/iclr/JainHGLYZWSSS25} dataset. We conduct experiments with three widely used open-sourced LLMs: CodeLlama~\cite{codellama}, DeepSeek-Coder~\cite{deepseek}, and Qwen2.5-Coder~\cite{qwen}. More experimental details are provided in Appendix~\ref{app:setup}.

\subsection{Main Results}

As shown in Table~\ref{tab:main_result}, our proposed \model{} approach consistently and significantly outperforms all baseline decoding strategies across all LLMs and benchmarks, with average relative improvement ranging from 16.9\% to 34.6\%. These results highlight the general effectiveness and broad applicability of \model{} in enhancing user-intent compliance in code generation.

The performance gain is most notable on our \mbox{\data} benchmark and IFEvalCode benchmark, which are explicitly designed to assess the constraint-following ability of
LLMs. Across models and datasets, \model{} achieves consistent relative improvements ranging from 30.5\% to 71.0\%, highlighting its effectiveness in improving models' constraint-following ability.

\model{} not only improves models' constraint-following ability, but also obtains clear gains on general code generation benchmarks such as HumanEval and LiveCodeBench. These results indicate that \model{} enhances constraint compliance while preserving and often improving the ability to generate functionally correct code.

Finally, compared to Selective Prompt Anchoring, \model{} does not require grid search tuning of an anchoring strength hyperparameter, thereby avoiding extra validation data and compute overhead. Across our experiments, \model{} yields larger and more stable gains. Moreover, while beam search can degrade generation quality~\cite{topp}, \model{} remains robust and still yields clear improvements when combined with beam search.

\subsection{Comparing Different Beam Sizes}

\begin{figure}
    \centering
    \includegraphics[width=0.99\linewidth]{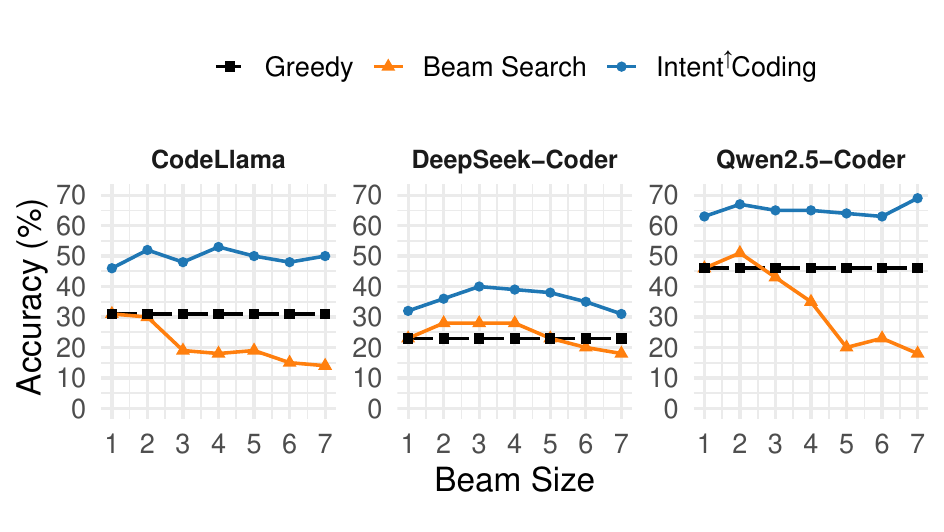}
    \caption{Performance with different beam sizes.}
    \label{fig:beam}
\end{figure}

Since \model{} is integrated with the beam search decoding process, we analyze its performance across different beam sizes. We conducted the analysis on our constructed \data{} dataset. The results in Figure~\ref{fig:beam} demonstrate that our \model{} achieves significant performance gains even with small beam sizes.
Notably, in the $beam\_size=1$ setting where our approach becomes an intent-amplified greedy search, \model{} still significantly outperforms greedy search across all LLMs. For instance, when applied to Qwen2.5-Coder, \model{} with a beam size of one achieves 63\% accuracy, a large improvement over the 46\% accuracy from the greedy search. While increasing the beam size in standard beam search yields only modest improvements and may even degrade performance at larger settings, which is consistent with prior findings~\cite{topp}, \model{} maintains high and stable performance at larger settings. This indicates that the performance gains arise not merely from broader search, but primarily from the effectiveness of the intent amplification mechanism in our approach. 

\subsection{Comparing Our Approach with a Fixed $\alpha$}
\begin{figure}
    \centering
    \includegraphics[width=0.99\linewidth]{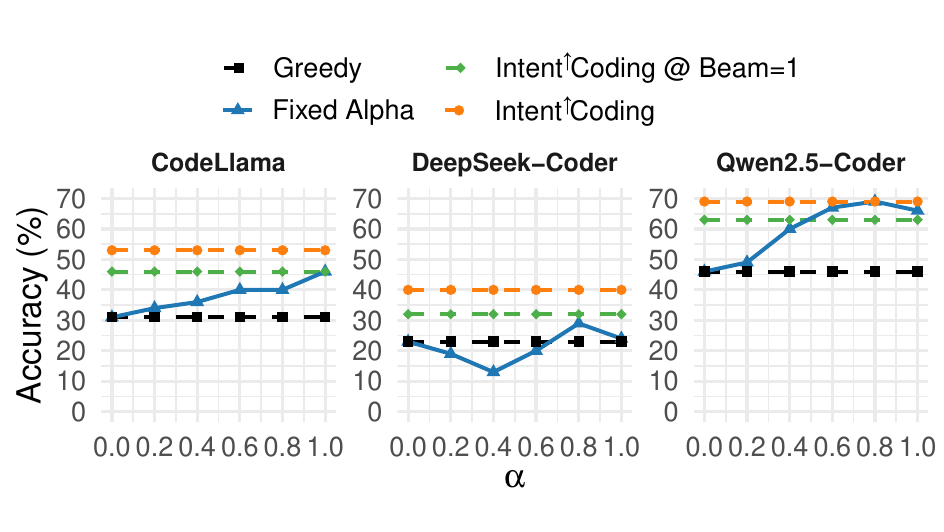}
    \caption{Comparing our ensemble of different $\alpha$ values with a fixed $\alpha$.}
    \label{fig:alpha}
\end{figure}

Our approach dynamically adjusts the contrastive strength by building an ensemble using different $\alpha$ values in Eqn.~\eqref{eq:alpha}.
To validate the effectiveness of such a strategy, we compare \model{} with a simpler baseline that uses a fixed $\alpha$, such as SPA. As shown in Figure~\ref{fig:alpha},  a fixed alpha baseline is highly sensitive to the choice of hyperparameter $\alpha$, and the optimal $\alpha$ value varies significantly across different LLMs. For example, CodeLlama performs best at $\alpha=1.0$, whereas Qwen2.5-Coder achieves its peak at $\alpha=0.8$. Moreover, the performance gap between different $\alpha$ values within the same LLM can be substantial. This sensitivity necessitates an expensive, model-specific hyperparameter search, reducing the practicality and generalizability of the fixed-$\alpha$ approach. By contrast, our \model{} employs a predefined set of $\alpha$ values and a token-level ensemble method. Thus, it does not require manual tuning, yet consistently outperforms the best $\alpha$ for each LLM. Notably, even under the $beam\_size=1$ setting, \model{} outperforms the fixed-$\alpha$ baseline in almost all cases. These results demonstrate that our dynamic scaling and ensemble strategy are robust and adaptable in amplifying user intent during code generation.

\subsection{LLMs of Different Mask Settings}
\begin{figure}[tb]        
  \centering
  \includegraphics[width=\columnwidth]{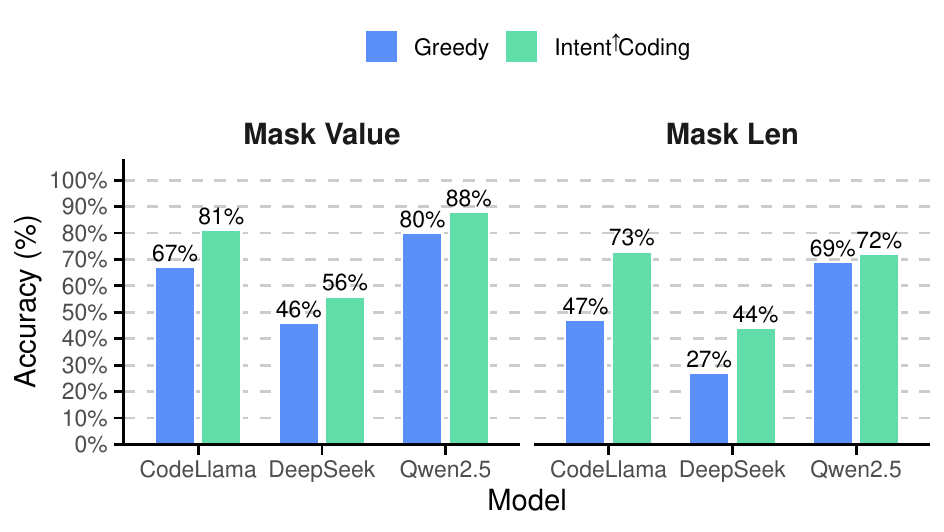}
  \caption{Performance of different constraint-mask settings. Left: Masking the constraint of variable values; Right: Masking the constraint of list/tuple/set length. }
  \label{fig:mask}
\end{figure}

In our standard \model{}, the intent-masked prompt is constructed by masking the entire user intent (containing all constraints). However, our approach is flexible and supports a more fine-grained variant, where only one or a subset of constraints is masked, allowing the amplification of the specific constraint(s). To assess this capability, we conducted experiments on the \data{} benchmark by masking an individual constraint. As shown in Figure~\ref{fig:mask}, \model{} consistently improves performance on the targeted constraint across all LLMs and constraint types. For example, when amplifying only the length constraint, all LLMs demonstrate a substantial boost in compliance of the length constraint compared with greedy decoding. 
The result demonstrates the granular control by our approach: \model{} offers a robust and adaptable mechanism for enhancing user intent, not only holistically, but also with precision for a specific constraint in code generation.

\subsection{Performance with Different LLMs}
\paragraph{Instruction-Tuned Models.}

Instruction-tuning is a widely adopted technique to improve LLMs’ alignment with user intent, and instruction-tuned models generally demonstrate stronger compliance with user-specified constraints~\cite{DBLP:journals/corr/abs-2304-03277}. To assess the effectiveness of \model{} on such LLMs, we conducted experiments using the instruction-tuned versions of CodeLlama, DeepSeek-Coder, and Qwen2.5-Coder. 
As shown in Table~\ref{tab:instruction}, \model{} provides significant performance gains even in this stronger setting.
It boosts the accuracy of CodeLlama-IT from  31\% to 53\% and DeepSeek-Code-IT from 52\% to 69\%, consistently outperforming the best results from optimized greedy or beam search baselines. These results suggest that \model{} is complementary to instruction-tuning, providing a further layer of capturing and enhancing user intent.


\begin{table}[tb]
  \centering
  \begin{tabular}{lcc}
    \toprule
    \textbf{Model} & \textbf{Baseline} & \textbf{\model{}} \\
    \midrule
    CodeLlama‑IT        & 31.0  & 53.0$_{\text{+71.0}\%}$ \\
    DeepSeek‑Coder‑IT   & 52.0  & 69.0$_{\text{+32.7}\%}$ \\
    Qwen2.5-Coder‑IT & 56.0  & 63.0$_{\text{+12.5}\%}$ \\
    \bottomrule
  \end{tabular}
  \caption{Performance with instruction-tuned models: CodeLlama-7b-Instruct-hf, DeepSeek-Coder-6.7b-Instruct, Qwen2.5-Coder-7b-Instruct. More experimental settings are provided in Appendix~\ref{app:dsetting}. }
  \label{tab:instruction}
\end{table}

\paragraph{Performance with Different Model Sizes.}

\begin{table}[tb]
  \centering
  \begin{tabular}{l@{\hskip 8pt}c@{\hskip 8pt}c}
    \toprule
    \textbf{Model} & \textbf{Baseline} & \textbf{\model{}} \\
    \midrule
    \textbf{CodeLlama} & & \\
    \quad 13B & 31.0 & 39.0$_{\text{+25.8}\%}$ \\
    \quad 34B & 35.0 & 46.0$_{\text{+31.4}\%}$ \\
    \midrule
    \textbf{DeepSeek-Coder} & & \\
    \quad 33B & 40.0 & 41.0$_{\text{+2.5}\%}$ \\
    \midrule
    \textbf{Qwen2.5-Coder} & & \\
    \quad 1.5B & 11.0 & 17.0$_{\text{+54.5}\%}$ \\
    \quad 3B   & 19.0 & 34.0$_{\text{+78.9}\%}$ \\
    \quad 14B  & 48.0 & 63.0$_{\text{+31.3}\%}$ \\
    \quad 32B  & 49.0 & 66.0$_{\text{+34.7}\%}$ \\
    \bottomrule
  \end{tabular}
  \caption{Performance with different model sizes on the \data{} dataset. More experimental settings are provided in Appendix~\ref{app:dsetting}.}
  \label{tab:size}
\end{table}

To assess the scalability and generalizability of our approach, we evaluated \model{} on a diverse set of LLMs ranging from 1.5B to 33B. 
As shown in Table~\ref{tab:size}, \model{} consistently outperforms the standard decoding baseline across all model sizes. For example, it improves the performance of the smaller Qwen2.5-Coder‑1.5B model from 11\% to 17\% (a 54.5\% relative improvement), while yielding a 17-point absolute gain for the much larger Qwen2.5-Coder‑32B model (a 34.7\% relative improvement). These results demonstrate that the benefits of \model{} hold across both lightweight and large-scale LLMs, confirming its robustness and model-agnostic nature.

\section{Conclusion} \label{sec:conclusion}
In this work, we proposed Intent-Amplified Code Generation (\model), a decoding strategy that improves LLMs' ability to follow user intent in multi-constraint code generation tasks. By extracting and amplifying the influence of user intent through a multi-strength ensemble mechanism, our approach enhances constraint satisfaction without requiring additional training or model modification. To support systematic evaluation, we introduced \data{}, a benchmark dataset designed to test LLMs under varying levels of constraint complexity. Extensive experiments across multiple LLMs and datasets demonstrate that \model{} consistently improves both functional correctness and intent compliance, highlighting its generality and effectiveness.


\section*{Limitations}
We discuss several limitations of our work.

First, \mbox{\data} covers four constraint types, which are relatively simple; while these constraints are common in practice, this design limits the benchmark’s coverage of more complex real-world requirements. In future work, we will extend \mbox{\data} to include a broader range of constraints, such as stylistic, architectural, and repository-specific constraints observed in real-world repositories.

Second, \model{} requires batching the original prompt with an intent-masked variant, which introduces additional compute overhead. As analyzed in Appendix~\ref{app:efficiency}, this overhead is modest in practice.

Third, \model{} is model-agnostic and can be applied to any LLM as long as token-level logits are available during decoding. However, popular closed-source APIs do not expose token-level logits or provide only limited information, which makes direct deployment difficult. Therefore, our experiments primarily focus on open-source LLMs.

\bibliography{custom}

@inproceedings{spa,
  author       = {Yuan Tian and
                  Tianyi Zhang},
  title        = {Selective Prompt Anchoring for Code Generation},
  booktitle    = {{ICML}},
  publisher    = {OpenReview.net},
  year         = {2025}
}

@article{ifevalcode,
  title={IFEvalCode: Controlled Code Generation},
  author={Yang, Jian and Zhang, Wei and Liu, Shukai and Chai, Linzheng and Tan, Yingshui and Liu, Jiaheng and Zhang, Ge and Zhou, Wangchunshu and Niu, Guanglin and Li, Zhoujun and others},
  journal={arXiv preprint arXiv:2507.22462},
  year={2025}
}

@article{li2022competition,
  title={Competition-level code generation with alphacode},
  author={Li, Yujia and Choi, David and Chung, Junyoung and Kushman, Nate and Schrittwieser, Julian and Leblond, R{\'e}mi and Eccles, Tom and Keeling, James and Gimeno, Felix and Dal Lago, Agustin and others},
  journal={Science},
  volume={378},
  number={6624},
  pages={1092--1097},
  year={2022}
}

@inproceedings{DBLP:conf/sigsoft/SvyatkovskiyDFS20,
  author       = {Alexey Svyatkovskiy and
                  Shao Kun Deng and
                  Shengyu Fu and
                  Neel Sundaresan},
  title        = {IntelliCode compose: code generation using transformer},
  booktitle    = {{ESEC/SIGSOFT} {FSE}},
  pages        = {1433--1443},
  publisher    = {{ACM}},
  year         = {2020}
}

@article{DBLP:journals/corr/abs-2107-03374,
  author       = {Mark Chen and
                  Jerry Tworek and
                  Heewoo Jun and
                  Qiming Yuan and
                  Henrique Pond{\'{e}} de Oliveira Pinto and
                  Jared Kaplan and
                  Harri Edwards and
                  Yuri Burda and
                  Nicholas Joseph and
                  Greg Brockman and
                  Alex Ray and
                  Raul Puri and
                  Gretchen Krueger and
                  Michael Petrov and
                  Heidy Khlaaf and
                  Girish Sastry and
                  Pamela Mishkin and
                  Brooke Chan and
                  Scott Gray and
                  Nick Ryder and
                  Mikhail Pavlov and
                  Alethea Power and
                  Lukasz Kaiser and
                  Mohammad Bavarian and
                  Clemens Winter and
                  Philippe Tillet and
                  Felipe Petroski Such and
                  Dave Cummings and
                  Matthias Plappert and
                  Fotios Chantzis and
                  Elizabeth Barnes and
                  Ariel Herbert{-}Voss and
                  William Hebgen Guss and
                  Alex Nichol and
                  Alex Paino and
                  Nikolas Tezak and
                  Jie Tang and
                  Igor Babuschkin and
                  Suchir Balaji and
                  Shantanu Jain and
                  William Saunders and
                  Christopher Hesse and
                  Andrew N. Carr and
                  Jan Leike and
                  Joshua Achiam and
                  Vedant Misra and
                  Evan Morikawa and
                  Alec Radford and
                  Matthew Knight and
                  Miles Brundage and
                  Mira Murati and
                  Katie Mayer and
                  Peter Welinder and
                  Bob McGrew and
                  Dario Amodei and
                  Sam McCandlish and
                  Ilya Sutskever and
                  Wojciech Zaremba},
  title        = {Evaluating Large Language Models Trained on Code},
  journal={arXiv preprint arXiv:2107.03374},
  year={2021}
}

@article{DBLP:journals/tosem/DongJJL24,
  author       = {Yihong Dong and
                  Xue Jiang and
                  Zhi Jin and
                  Ge Li},
  title        = {Self-Collaboration Code Generation via ChatGPT},
  journal      = {TOSEM},
  volume       = {33},
  number       = {7},
  pages        = {189:1--189:38},
  year         = {2024}
}

@article{DBLP:journals/corr/abs-2406-00515,
  author       = {Juyong Jiang and
                  Fan Wang and
                  Jiasi Shen and
                  Sungju Kim and
                  Sunghun Kim},
  title        = {A Survey on Large Language Models for Code Generation},
  journal={arXiv preprint arXiv:2406.00515},
  year={2024}
}

@article{codellama,
  author       = {Baptiste Rozi{\`{e}}re and
                  Jonas Gehring and
                  Fabian Gloeckle and
                  Sten Sootla and
                  Itai Gat and
                  Xiaoqing Ellen Tan and
                  Yossi Adi and
                  Jingyu Liu and
                  Tal Remez and
                  J{\'{e}}r{\'{e}}my Rapin and
                  Artyom Kozhevnikov and
                  Ivan Evtimov and
                  Joanna Bitton and
                  Manish Bhatt and
                  Cristian Canton{-}Ferrer and
                  Aaron Grattafiori and
                  Wenhan Xiong and
                  Alexandre D{\'{e}}fossez and
                  Jade Copet and
                  Faisal Azhar and
                  Hugo Touvron and
                  Louis Martin and
                  Nicolas Usunier and
                  Thomas Scialom and
                  Gabriel Synnaeve},
  title        = {Code Llama: Open Foundation Models for Code},
  journal={arXiv preprint arXiv:2308.12950},
  year={2023}
}

@article{DBLP:journals/tosem/JiangDWFSLJJ24,
  author       = {Xue Jiang and
                  Yihong Dong and
                  Lecheng Wang and
                  Zheng Fang and
                  Qiwei Shang and
                  Ge Li and
                  Zhi Jin and
                  Wenpin Jiao},
  title        = {Self-Planning Code Generation with Large Language Models},
  journal      = {TOSEM},
  volume       = {33},
  number       = {7},
  pages        = {182:1--182:30},
  year         = {2024}
}

@inproceedings{DBLP:conf/iclr/NijkampPHTWZSX23,
  author       = {Erik Nijkamp and
                  Bo Pang and
                  Hiroaki Hayashi and
                  Lifu Tu and
                  Huan Wang and
                  Yingbo Zhou and
                  Silvio Savarese and
                  Caiming Xiong},
  title        = {CodeGen: An Open Large Language Model for Code with Multi-Turn Program
                  Synthesis},
  booktitle    = {{ICLR}},
  year         = {2023}
}

@inproceedings{DBLP:conf/aaai/ZhuLLZLJ024,
  author       = {Yuqi Zhu and
                  Jia Li and
                  Ge Li and
                  Yunfei Zhao and
                  Jia Li and
                  Zhi Jin and
                  Hong Mei},
  title        = {Hot or Cold? Adaptive Temperature Sampling for Code Generation with
                  Large Language Models},
  booktitle    = {{AAAI}},
  pages        = {437--445},
  year         = {2024}
}

@inproceedings{DBLP:conf/aclnmt/FreitagA17,
  author       = {Markus Freitag and
                  Yaser Al{-}Onaizan},
  title        = {Beam Search Strategies for Neural Machine Translation},
  booktitle    = {{ACL}},
  pages        = {56--60},
  year         = {2017}
}

@inproceedings{DBLP:conf/acl/LiHFLEHZL23,
  author       = {Xiang Lisa Li and
                  Ari Holtzman and
                  Daniel Fried and
                  Percy Liang and
                  Jason Eisner and
                  Tatsunori Hashimoto and
                  Luke Zettlemoyer and
                  Mike Lewis},
  title        = {Contrastive Decoding: Open-ended Text Generation as Optimization},
  booktitle    = {{ACL}},
  pages        = {12286--12312},
  year         = {2023}
}

@inproceedings{DBLP:conf/iclr/KimKLY24,
  author       = {Taehyeon Kim and
                  Joonkee Kim and
                  Gihun Lee and
                  Se{-}Young Yun},
  title        = {Instructive Decoding: Instruction-Tuned Large Language Models are
                  Self-Refiner from Noisy Instructions},
  booktitle    = {{ICLR}},
  year         = {2024}
}

@inproceedings{DBLP:conf/iclr/JainHGLYZWSSS25,
  author       = {Naman Jain and
                  King Han and
                  Alex Gu and
                  Wen{-}Ding Li and
                  Fanjia Yan and
                  Tianjun Zhang and
                  Sida Wang and
                  Armando Solar{-}Lezama and
                  Koushik Sen and
                  Ion Stoica},
  title        = {LiveCodeBench: Holistic and Contamination Free Evaluation of Large
                  Language Models for Code},
  booktitle    = {{ICLR}},
  year         = {2025}
}

@article{starcoder,
  author       = {Raymond Li and
                  Loubna Ben Allal and
                  Yangtian Zi and
                  Niklas Muennighoff and
                  Denis Kocetkov and
                  Chenghao Mou and
                  Marc Marone and
                  Christopher Akiki and
                  Jia Li and
                  Jenny Chim and
                  Qian Liu and
                  Evgenii Zheltonozhskii and
                  Terry Yue Zhuo and
                  Thomas Wang and
                  Olivier Dehaene and
                  Mishig Davaadorj and
                  Joel Lamy{-}Poirier and
                  Jo{\~{a}}o Monteiro and
                  Oleh Shliazhko and
                  Nicolas Gontier and
                  Nicholas Meade and
                  Armel Zebaze and
                  Ming{-}Ho Yee and
                  Logesh Kumar Umapathi and
                  Jian Zhu and
                  Benjamin Lipkin and
                  Muhtasham Oblokulov and
                  Zhiruo Wang and
                  Rudra Murthy V and
                  Jason T. Stillerman and
                  Siva Sankalp Patel and
                  Dmitry Abulkhanov and
                  Marco Zocca and
                  Manan Dey and
                  Zhihan Zhang and
                  Nour Fahmy and
                  Urvashi Bhattacharyya and
                  Wenhao Yu and
                  Swayam Singh and
                  Sasha Luccioni and
                  Paulo Villegas and
                  Maxim Kunakov and
                  Fedor Zhdanov and
                  Manuel Romero and
                  Tony Lee and
                  Nadav Timor and
                  Jennifer Ding and
                  Claire Schlesinger and
                  Hailey Schoelkopf and
                  Jan Ebert and
                  Tri Dao and
                  Mayank Mishra and
                  Alex Gu and
                  Jennifer Robinson and
                  Carolyn Jane Anderson and
                  Brendan Dolan{-}Gavitt and
                  Danish Contractor and
                  Siva Reddy and
                  Daniel Fried and
                  Dzmitry Bahdanau and
                  Yacine Jernite and
                  Carlos Mu{\~{n}}oz Ferrandis and
                  Sean Hughes and
                  Thomas Wolf and
                  Arjun Guha and
                  Leandro von Werra and
                  Harm de Vries},
  title        = {StarCoder: may the source be with you!},
  journal      = {TMLR},
  volume       = {2023},
  year         = {2023}
}

@article{deepseek,
  author       = {Daya Guo and
                  Qihao Zhu and
                  Dejian Yang and
                  Zhenda Xie and
                  Kai Dong and
                  Wentao Zhang and
                  Guanting Chen and
                  Xiao Bi and
                  Y. Wu and
                  Y. K. Li and
                  Fuli Luo and
                  Yingfei Xiong and
                  Wenfeng Liang},
  title        = {DeepSeek-Coder: When the Large Language Model Meets Programming -
                  The Rise of Code Intelligence},
  journal={arXiv preprint arXiv:2401.14196},
  year={2024}
}

@article{qwen,
  author       = {Binyuan Hui and
                  Jian Yang and
                  Zeyu Cui and
                  Jiaxi Yang and
                  Dayiheng Liu and
                  Lei Zhang and
                  Tianyu Liu and
                  Jiajun Zhang and
                  Bowen Yu and
                  Kai Dang and
                  An Yang and
                  Rui Men and
                  Fei Huang and
                  Xingzhang Ren and
                  Xuancheng Ren and
                  Jingren Zhou and
                  Junyang Lin},
  title        = {Qwen2.5-Coder Technical Report},
  journal={arXiv preprint arXiv:2409.12186},
  year={2024}
}

@inproceedings{self-debug,
  author       = {Xinyun Chen and
                  Maxwell Lin and
                  Nathanael Sch{\"{a}}rli and
                  Denny Zhou},
  title        = {Teaching Large Language Models to Self-Debug},
  booktitle    = {{ICLR}},
  year         = {2024}
}

@inproceedings{DBLP:conf/iclr/ZhangCSDTG23,
  author       = {Shun Zhang and
                  Zhenfang Chen and
                  Yikang Shen and
                  Mingyu Ding and
                  Joshua B. Tenenbaum and
                  Chuang Gan},
  title        = {Planning with Large Language Models for Code Generation},
  booktitle    = {{ICLR}},
  year         = {2023}
}

@inproceedings{DBLP:conf/acl/ZhangLLLJ23,
  author       = {Kechi Zhang and
                  Zhuo Li and
                  Jia Li and
                  Ge Li and
                  Zhi Jin},
  title        = {Self-Edit: Fault-Aware Code Editor for Code Generation},
  booktitle    = {{ACL}},
  pages        = {769--787},
  year         = {2023}
}

@inproceedings{rocode,
  author       = {Xue Jiang and
                  Yihong Dong and
                  Yongding Tao and
                  Huanyu Liu and
                  Zhi Jin and
                  Ge Li},
  title        = {{ROCODE:} Integrating Backtracking Mechanism and Program Analysis
                  in Large Language Models for Code Generation},
  booktitle    = {{ICSE}},
  pages        = {334--346},
  year         = {2025}
}

@article{mbpp,
  author       = {Jacob Austin and
                  Augustus Odena and
                  Maxwell I. Nye and
                  Maarten Bosma and
                  Henryk Michalewski and
                  David Dohan and
                  Ellen Jiang and
                  Carrie J. Cai and
                  Michael Terry and
                  Quoc V. Le and
                  Charles Sutton},
  title        = {Program Synthesis with Large Language Models},
  journal={arXiv preprint arXiv:2108.07732},
  year={2021}
}

@inproceedings{apps,
  author       = {Dan Hendrycks and
                  Steven Basart and
                  Saurav Kadavath and
                  Mantas Mazeika and
                  Akul Arora and
                  Ethan Guo and
                  Collin Burns and
                  Samir Puranik and
                  Horace He and
                  Dawn Song and
                  Jacob Steinhardt},
  title        = {Measuring Coding Challenge Competence With {APPS}},
  booktitle    = {NeurIPS Datasets and Benchmarks Track},
  year         = {2021}
}

@inproceedings{temperature,
  author       = {Matthew Renze},
  title        = {The Effect of Sampling Temperature on Problem Solving in Large Language
                  Models},
  booktitle    = {{EMNLP}},
  pages        = {7346--7356},
  year         = {2024}
}

@inproceedings{topk,
  author       = {Angela Fan and
                  Mike Lewis and
                  Yann N. Dauphin},
  title        = {Hierarchical Neural Story Generation},
  booktitle    = {{ACL}},
  pages        = {889--898},
  year         = {2018}
}

@inproceedings{topp,
  author       = {Ari Holtzman and
                  Jan Buys and
                  Li Du and
                  Maxwell Forbes and
                  Yejin Choi},
  title        = {The Curious Case of Neural Text Degeneration},
  booktitle    = {{ICLR}},
  year         = {2020}
}

@inproceedings{DBLP:conf/scam/SiddiqDSS24,
  author       = {Mohammed Latif Siddiq and
                  Simantika Dristi and
                  Joy Saha and
                  Joanna C. S. Santos},
  title        = {The Fault in our Stars: Quality Assessment of Code Generation Benchmarks},
  booktitle    = {{SCAM}},
  pages        = {201--212},
  year         = {2024}
}

@article{DBLP:journals/corr/abs-2405-11430,
  author       = {Jianbo Dai and
                  Jianqiao Lu and
                  Yunlong Feng and
                  Rongju Ruan and
                  Ming Cheng and
                  Haochen Tan and
                  Zhijiang Guo},
  title        = {{MHPP:} Exploring the Capabilities and Limitations of Language Models
                  Beyond Basic Code Generation},
  journal={arXiv preprint arXiv:2405.11430},
  year={2024}
}

@inproceedings{DBLP:conf/emnlp/ShiY0ZWYL24,
  author       = {Chufan Shi and
                  Haoran Yang and
                  Deng Cai and
                  Zhisong Zhang and
                  Yifan Wang and
                  Yujiu Yang and
                  Wai Lam},
  title        = {A Thorough Examination of Decoding Methods in the Era of LLMs},
  booktitle    = {{EMNLP}},
  pages        = {8601--8629},
  year         = {2024}
}

@inproceedings{DBLP:conf/acl/IppolitoKSKC19,
  author       = {Daphne Ippolito and
                  Reno Kriz and
                  Jo{\~{a}}o Sedoc and
                  Maria Kustikova and
                  Chris Callison{-}Burch},
  title        = {Comparison of Diverse Decoding Methods from Conditional Language Models},
  booktitle    = {{ACL}},
  pages        = {3752--3762},
  year         = {2019}
}

@article{llama2,
  author       = {Hugo Touvron and
                  Louis Martin and
                  Kevin Stone and
                  Peter Albert and
                  Amjad Almahairi and
                  Yasmine Babaei and
                  Nikolay Bashlykov and
                  Soumya Batra and
                  Prajjwal Bhargava and
                  Shruti Bhosale and
                  Dan Bikel and
                  Lukas Blecher and
                  Cristian Canton{-}Ferrer and
                  Moya Chen and
                  Guillem Cucurull and
                  David Esiobu and
                  Jude Fernandes and
                  Jeremy Fu and
                  Wenyin Fu and
                  Brian Fuller and
                  Cynthia Gao and
                  Vedanuj Goswami and
                  Naman Goyal and
                  Anthony Hartshorn and
                  Saghar Hosseini and
                  Rui Hou and
                  Hakan Inan and
                  Marcin Kardas and
                  Viktor Kerkez and
                  Madian Khabsa and
                  Isabel Kloumann and
                  Artem Korenev and
                  Punit Singh Koura and
                  Marie{-}Anne Lachaux and
                  Thibaut Lavril and
                  Jenya Lee and
                  Diana Liskovich and
                  Yinghai Lu and
                  Yuning Mao and
                  Xavier Martinet and
                  Todor Mihaylov and
                  Pushkar Mishra and
                  Igor Molybog and
                  Yixin Nie and
                  Andrew Poulton and
                  Jeremy Reizenstein and
                  Rashi Rungta and
                  Kalyan Saladi and
                  Alan Schelten and
                  Ruan Silva and
                  Eric Michael Smith and
                  Ranjan Subramanian and
                  Xiaoqing Ellen Tan and
                  Binh Tang and
                  Ross Taylor and
                  Adina Williams and
                  Jian Xiang Kuan and
                  Puxin Xu and
                  Zheng Yan and
                  Iliyan Zarov and
                  Yuchen Zhang and
                  Angela Fan and
                  Melanie Kambadur and
                  Sharan Narang and
                  Aur{\'{e}}lien Rodriguez and
                  Robert Stojnic and
                  Sergey Edunov and
                  Thomas Scialom},
  title        = {Llama 2: Open Foundation and Fine-Tuned Chat Models},
  journal={arXiv preprint arXiv:2307.09288},
  year={2023}
}

@article{DBLP:journals/corr/abs-2304-03277,
  author       = {Baolin Peng and
                  Chunyuan Li and
                  Pengcheng He and
                  Michel Galley and
                  Jianfeng Gao},
  title        = {Instruction Tuning with {GPT-4}},
  journal      = {arXiv preprint arXiv:2304.03277},
  year         = {2023}
}

\appendix

\clearpage

\setcounter{figure}{0}
\setcounter{table}{0}

\section{Experimental Details}
\label{app:setup}
\subsection{Datasets}

\textbf{HumanEval}~\cite{DBLP:journals/corr/abs-2107-03374} is a widely-used benchmark dataset for evaluating code generation capabilities of LLMs. It consists of 164 hand-crafted Python programming problems, each defined by a function signature and a natural language description. 

\noindent \textbf{LiveCodeBench}~\cite{DBLP:conf/iclr/JainHGLYZWSSS25} offers a more realistic evaluation of code generation capabilities by sourcing problems directly from programming contests. These tasks are characterized by complex, natural language descriptions and are evaluated against a set of test cases. In our experiments, we utilize Version 5 of the dataset, which comprises 279 easy, 331 medium, and 270 hard problems.

\noindent
\textbf{IFEvalCode}~\cite{ifevalcode} pairs each code generation task with explicit, verifiable instruction constraints and an executable checker that tests whether the generated code satisfies those constraints. In our experiments, we report the instruction compliance metric (Instr.) proposed by this benchmark on its python and english questions. 

In addition to existing datasets, our work also constructed a new dataset,  \textbf{\mbox{\data}}, to evaluate LLMs' compliance with user-specified constraints (detailed in the previous section). In our experiments, we focus on Level 4, which is the most challenging setting, having four constraints in each data sample.

\subsection{Base Models}
We conducted experiments with popular open-source LLMs, as we need LLM logit during decoding. In particular, we consider the following models.

\noindent \textbf{CodeLlama}~\cite{codellama} is an open-source family of code-specialized LLMs, based on the LLaMa model~\cite{llama2} and further pre-trained on permissively licensed code. It has demonstrated strong performance on standard code generation benchmarks. In our experiments, we used the CodeLlama‑7B‑hf variant.

\noindent \textbf{DeepSeek-Coder}~\cite{deepseek} is a family of open-source LLMs trained from scratch on 2 trillion tokens, comprising 87\% code and 13\% natural language in both English and Chinese. The LLMs are pre-trained on a project-level code corpus with a 16K context window and an additional fill-in-the-blank objective to support long-range code completion and infilling. In our experiments, we used the DeepSeek-Coder‑6.7B‑Base model.

\noindent \textbf{Qwen2.5-Coder}~\cite{qwen} is a series of code-specific LLMs, extending the Qwen2.5 model with specialized training on 5.5 trillion tokens. It supports long-context understanding up to 128K tokens and exhibits strong performance in code generation, reasoning, and fixing. In our experiments, we used the Qwen2.5-Coder‑7B model.

\subsection{Implementation Details}
In our experiment, we explored multiple decoding strategies as baselines to ensure a comprehensive comparison. For both standard beam search and our \model{} method, the beam size was set to 4. In nucleus sampling~\cite{topp}, we used top-$p$ with a $p$ value chosen from \{0.7, 0.8, 0.9\} and a fixed temperature of 0.2. For the AdapT sampling method~\cite{DBLP:conf/aaai/ZhuLLZLJ024}, we fixed top-p at 0.95, and performed a grid search over its temperature parameters $T_1$ and $T_2$; we report the results from the best configuration. All sampling-based methods are run three times, and we report the mean and standard deviation. For the Select Prompt Anchoring method, we performed a grid search over its anchoring strength $\omega$; we report the results from the best configuration. The maximum generation length is set to 512 for HumanEval and IFEvalCode, 1024 for LiveCodeBench, and 256 for our \data{} dataset. For the HumanEval and LiveCodeBench datasets, we report results in the standard $\mathrm{pass@1}$ metric, whereas for the CodeConstraints benchmark, we use the accuracy metric, where the code is considered accurate if all constraints are satisfied.

\subsection{Performance with Different LLMs Settings}
\label{app:dsetting}
Baseline denotes the best performance achieved by either greedy search or beam search. For both methods, we report the best performance from a grid search over beam sizes \{1, 2, 3, 4\}. The subscript values represent the relative improvement of \model{} over the baseline.

\section{Inference Efficiency}
\label{app:efficiency}
We analyze the inference efficiency of our \model{} approach. We considered the \data{} dataset and used a single NVIDIA A100‑40G GPU. For a fair comparison, all methods, greedy, beam search, and \model{}, were run with the \texttt{early\_stop="never"} configuration, ensuring that generation continues until all beams produce an EOS token. Each experiment was repeated five times to reduce measurement variance, and we report the average and standard deviation  in Table~\ref{tab:data}.

Results show that our \model{} is slightly slower than greedy and beam search in terms of time per token. This is expected because our approach involves an ensemble of multiple contrastive strengths. However, our method often outputs fewer tokens than beam search (which tends to generate code snippets that are less related to the user intent); consequently, our approach is more efficient than beam search in terms of inference time. Overall, our \mbox{\model} is an efficient decoding method for code generation.

\section{Case Studies}
Table~\ref{tab:case} presents an example involving four user-specified constraints: (1) The code should return a set, (2) The element in set should be integer, (3) The integer in set should be greater than or equal to $n$, and (4) The length of the set should be 99. 

The code generated by greedy decoding satisfies the first two constraints but fails to ensure the correct value and set size. Beam search decoding improves upon this by correctly constraining the value, but it still fails to guarantee the exact set length. Additionally, it produces unrelated code after finishing the function body. In contrast, our \mbox{\model{}} approach generates code that fully satisfies all four constraints and terminates precisely at the end of the intended logic, without producing additional tokens. Overall, this example highlights the strength and efficiency of \mbox{\model{}} in capturing fine-grained constraints. 

\begin{table*}[t]
\centering
\small

\begin{tabular}{llccc}
\toprule
\textbf{Model} & \textbf{Method} & \textbf{Avg. Time (s)} & \textbf{Avg. Tokens} & \textbf{Time/Token (ms)} \\
\midrule

\multirow{4}{*}{CodeLlama}
& Greedy         & 7.84$\pm0.04$  & 172.48$\pm0.00$ & 45.30$\pm0.04$\\
& \model{}@1       & 10.45$\pm0.04$  & 211.62$\pm0.00$ & 49.34$\pm0.04$\\
& Beam Search@4    & 12.65$\pm0.06$ & 253.53$\pm0.00$ & 50.21$\pm0.21$\\
& \model{}@4 & 10.82$\pm0.05$ & 210.41$\pm0.00$ & 51.24$\pm0.11$\\

\midrule
\multirow{4}{*}{DeepSeek-Coder}
& Greedy         & 8.65$\pm0.04$  & 192.58$\pm0.00$ & 44.83$\pm0.03$ \\
& \model{}@1       & 10.12$\pm0.07$ & 206.47$\pm0.00$ & 49.04 $\pm0.06$ \\
& Beam Search@4    & 12.65$\pm0.09$ & 255.04$\pm0.00$ & 49.72$\pm0.14$\\
& \model{}@4 & 12.12$\pm0.04$ & 235.69$\pm0.00$ & 51.26$\pm0.08$\\

\midrule
\multirow{4}{*}{Qwen2.5-Coder}
& Greedy         & 8.61$\pm0.06$  & 185.51$\pm0.00$ & 45.86$\pm0.03$\\
& \model{}@1       & 5.13$\pm0.06$  & 104.44$\pm0.00$ & 49.09$\pm0.06$\\
& Beam Search@4    & 12.44$\pm0.04$ & 254.23$\pm0.00$ & 49.06$\pm0.13$\\
& \model{}@4 & 7.38$\pm0.05$  & 146.09$\pm0.00$ & 50.82$\pm0.16$\\

\bottomrule
\end{tabular}
\caption{Average inference time (in seconds), maximum generated token count per problem and average inference time per token (in milliseconds) on the \data{} benchmark, using a single NVIDIA A100‑40G GPU. Results are reported for 7B base models with beam  sizes of one (@1) and four (@4).}
\label{tab:data}
\vspace{4em}
\includegraphics[width=0.99\linewidth]{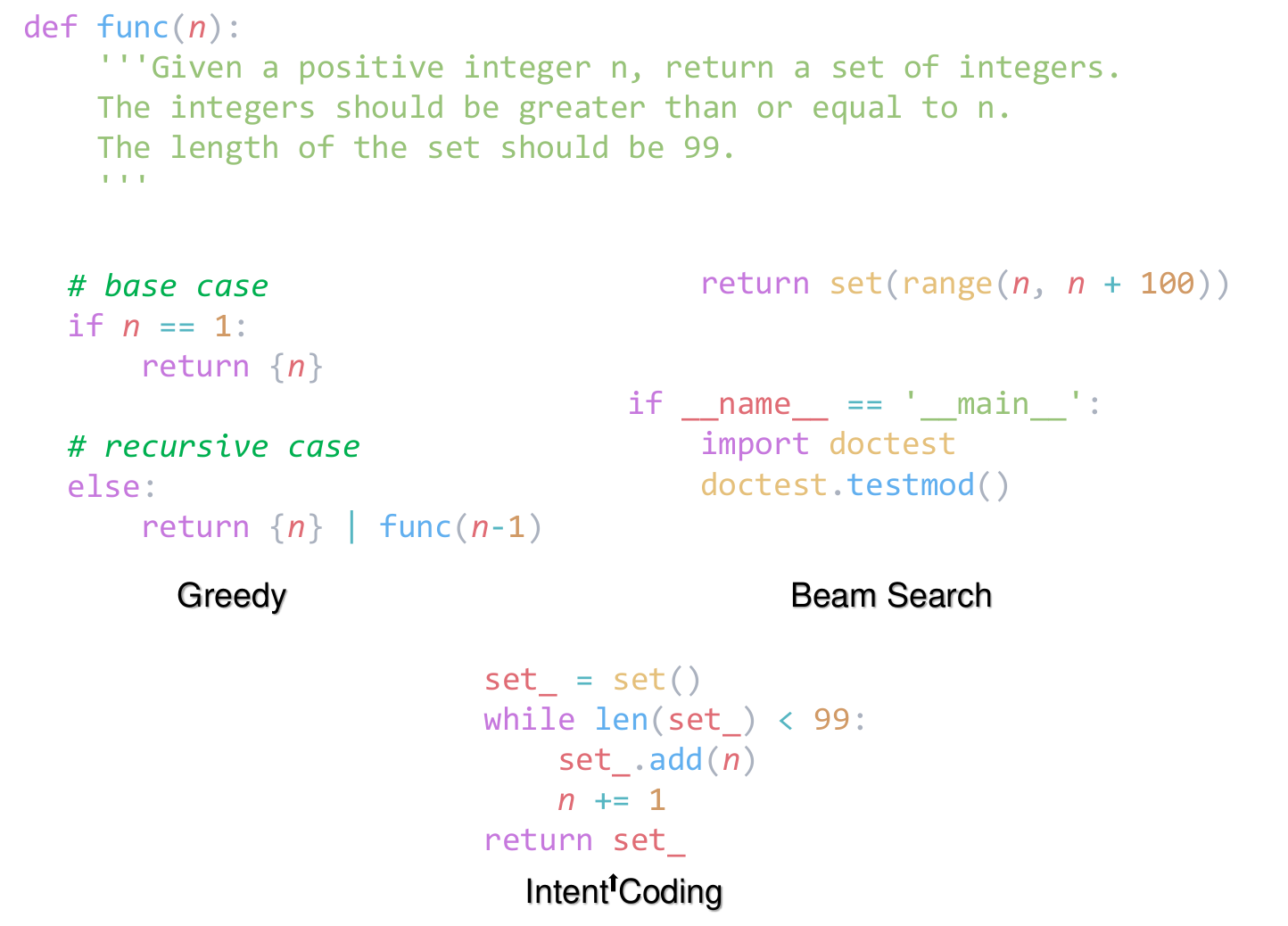}
\caption{ Case studies.}
\label{tab:case}
\end{table*}

\section{Token-Level Ensemble Details}
\label{app:tok}
As described in Section~\ref{para:tok}, at each decoding step we obtain six logit distributions $\{\tilde{o}^{(k)}\}_{k=1}^{6}$ and the corresponding probabilities
$p^{(k)}=\mathrm{softmax}(\tilde{o}^{(k)})$. For each $k$, we take the top-1 token
\begin{equation}
t^{(k)}=\arg\max_{v} p^{(k)}_{v},
\qquad
s^{(k)}=p^{(k)}_{t^{(k)}} .
\end{equation}
Since multiple $t^{(k)}$ may share the same token id $v$, we group candidates by id and
assign an ensemble score
\begin{equation}
\begin{aligned}
\bar{p}(v) &= \frac{1}{|S(v)|}\sum_{k\in S(v)} s^{(k)},\\
S(v) &= \{k \mid t^{(k)} = v\}.
\end{aligned}
\end{equation}
We return each unique token id $v$ with $\bar{p}(v)$. For example, if the six top-1 tokens
$T_1,\dots,T_6$ have probabilities $P_1,\dots,P_6$, where $T_1,T_2$ map to token id $A$
and $T_3,T_4,T_5,T_6$ map to token id $B$, then we return $A$ and $B$ with
$\bar{p}(A)=(P_1+P_2)/2$ and $\bar{p}(B)=(P_3+P_4+P_5+P_6)/4$.

\section{Prompt}
\label{app:prompt}
In \model{}, we construct an intent-masked prompt from each dataset-provided original prompt. We start from the raw prompt as released by the benchmark, then use Python regular expressions to automatically locate the natural-language intent span and mark it with a mask token while preserving the rest of the context. This masking pipeline is fully automated. The prompt templates for each dataset are shown below.
\subsection{HumanEval Prompt}
\lstset{
  basicstyle=\ttfamily\footnotesize,
  frame=single,
  breaklines=true,
  showstringspaces=false,
  keepspaces=true,
  columns=fullflexible,
  captionpos=b,
  keywordstyle=\relax,
  commentstyle=\relax,
  stringstyle=\relax,
}

\begin{lstlisting}[caption={HumanEval Original Prompt}, label={lst:ho}]
from typing import List


def has_close_elements(numbers: List[float], threshold: float) -> bool:
    """ Check if in given list of numbers, are any two numbers closer to each other than
    given threshold.
    >>> has_close_elements([1.0, 2.0, 3.0], 0.5)
    False
    >>> has_close_elements([1.0, 2.8, 3.0, 4.0, 5.0, 2.0], 0.3)
    True
    """
\end{lstlisting}

\begin{lstlisting}[caption={HumanEval Intent-Masked Prompt}, label={lst:hm}, literate={<mask>}{{\colorbox{black!10}{\texttt{<mask>}}}}6]
from typing import List


def has_close_elements(numbers: List[float], threshold: float) -> bool:
    <mask>
    """ Check if in given list of numbers, are any two numbers closer to each other than
    given threshold.
    >>> has_close_elements([1.0, 2.0, 3.0], 0.5)
    False
    >>> has_close_elements([1.0, 2.8, 3.0, 4.0, 5.0, 2.0], 0.3)
    True
    """
    <mask>
\end{lstlisting}

\subsection{LiveCodeBench Prompt}

\begin{lstlisting}[caption={LiveCodeBench Original Prompt}, label={lst:lo}]
### Question
You are given an integer sequence A=(A_1,\ldots,A_N) of length N. Here, A_1, A_2, \ldots, A_N are all distinct.
Which element in A is the second largest?

Input

The input is given from Standard Input in the following format:
N 
A_1 A_2 \ldots A_{N}

Output

Print the integer X such that the X-th element in A is the second largest.

Constraints


- 2 \leq N \leq 100
- 1 \leq A_i \leq 10^9
- A_1, A_2, \ldots, A_N are all distinct.
- All input values are integers.

Sample Input 1

4
8 2 5 1

Sample Output 1

3

The second largest element in A is A_3, so print 3.

Sample Input 2

8
1 2 3 4 5 10 9 11

Sample Output 2

6

### Answer
\end{lstlisting}

\begin{lstlisting}[caption={LiveCodeBench Intent-Masked Prompt}, label={lst:lm}, literate={<mask>}{{\colorbox{black!10}{\texttt{<mask>}}}}6]
### Question
<mask>
You are given an integer sequence A=(A_1,\ldots,A_N) of length N. Here, A_1, A_2, \ldots, A_N are all distinct.
Which element in A is the second largest?
<mask>

Input

The input is given from Standard Input in the following format:
N 
A_1 A_2 \ldots A_{N}

Output

Print the integer X such that the X-th element in A is the second largest.

Constraints


- 2 \leq N \leq 100
- 1 \leq A_i \leq 10^9
- A_1, A_2, \ldots, A_N are all distinct.
- All input values are integers.

Sample Input 1

4
8 2 5 1

Sample Output 1

3

The second largest element in A is A_3, so print 3.

Sample Input 2

8
1 2 3 4 5 10 9 11

Sample Output 2

6

### Answer
\end{lstlisting}

\subsection{\mbox{\data} Prompt}

\begin{lstlisting}[caption={\mbox{\data} Original Prompt}, label={lst:co}, literate={<mask>}{{\colorbox{black!10}{\texttt{<mask>}}}}6]
def func(n):
    '''
    Given a positive integer n, 
    return a tuple of floats. 
    The floats should be less than n. 
    The length of the tuple should be 2.
    '''
\end{lstlisting}

\begin{lstlisting}[caption={\mbox{\data} Intent-Masked Prompt}, label={lst:cm}, literate={<mask>}{{\colorbox{black!10}{\texttt{<mask>}}}}6]
def func(n):
    <mask>
    '''
    Given a positive integer n, 
    return a tuple of floats. 
    The floats should be less than n. 
    The length of the tuple should be 2.
    '''
    <mask>
\end{lstlisting}

\subsection{IFEvalCode Prompt}
\begin{lstlisting}[caption={IFEvalCode Original Prompt}, label={lst:io}, literate={<mask>}{{\colorbox{black!10}{\texttt{<mask>}}}}6]
Given a string, calculate the minimum cost to complete character escaping. The escaping rules are as follows:
- '<' escapes to '&lt;'
- '>' escapes to '&gt;'
- '&' escapes to '&amp;'
- '\"' escapes to '&quot;'
- '\'' escapes to '&#39;'

There are two types of escape operations:
1. Single character escape: Replace a single escapable character with its corresponding HTML entity, cost is 1 unit
2. Substring escape: Replace a continuous substring containing only escapable characters with their corresponding HTML entities, cost is 2 units (regardless of substring length)

Constraints:
1. Internal variable names must use snake_case
2. Must use list comprehension
3. Must use one ternary operator
4. Total code lines must not exceed 15 lines (including empty lines)
5. Must include at least one lambda expression

Function signature:```
def min_escape_cost(input_str: str) -> int:
    pass
```

Example input output:
```
Input: 'Hello <World> & \"Everyone\"'
Output: 5
Explanation: One optimal solution:
1. Escape '<' and '>' as substring, cost = 2
2. Escape '&' as single character, cost = 1
3. Escape two '\"' as substring, cost = 2
Total cost = 2 + 1 + 2 = 5
```

Please return all the complete code in the first code block.
\end{lstlisting}

\begin{lstlisting}[caption={IFEvalCode Intent-Masked Prompt}, label={lst:im}, literate={<mask>}{{\colorbox{black!10}{\texttt{<mask>}}}}6]
Given a string, calculate the minimum cost to complete character escaping. The escaping rules are as follows:
- '<' escapes to '&lt;'
- '>' escapes to '&gt;'
- '&' escapes to '&amp;'
- '\"' escapes to '&quot;'
- '\'' escapes to '&#39;'

There are two types of escape operations:
1. Single character escape: Replace a single escapable character with its corresponding HTML entity, cost is 1 unit
2. Substring escape: Replace a continuous substring containing only escapable characters with their corresponding HTML entities, cost is 2 units (regardless of substring length)

Constraints:
<mask>
1. Internal variable names must use snake_case
2. Must use list comprehension
3. Must use one ternary operator
4. Total code lines must not exceed 15 lines (including empty lines)
5. Must include at least one lambda expression
<mask>

Function signature:```
def min_escape_cost(input_str: str) -> int:
    pass
```

Example input output:
```
Input: 'Hello <World> & \"Everyone\"'
Output: 5
Explanation: One optimal solution:
1. Escape '<' and '>' as substring, cost = 2
2. Escape '&' as single character, cost = 1
3. Escape two '\"' as substring, cost = 2
Total cost = 2 + 1 + 2 = 5
```

Please return all the complete code in the first code block.
\end{lstlisting}

\end{document}